\documentclass[pra,twocolumn,aps]{revtex4}
\usepackage{graphicx}
\usepackage{float}
\usepackage{textcomp}
\usepackage{color}
\usepackage{amsmath}
\usepackage{psfrag,overpic,pst-all}
\usepackage{ulem}
\usepackage{enumitem}
\usepackage{siunitx}

\newcommand{\micron}{\textmu{m}}

\newcommand{\textmodcolor}{black}

\begin{document}

\title{Second Harmonic Generation Enabled by Longitudinal Electric Field Components in Photonic Wire Waveguides}

\author{Nicolas~Poulvellarie,$^{1,2,3}$ Utsav~Dave,$^{4}$ Koen~Alexander,$^{2,3}$ Charles~Ciret,$^{5}$  Maximilien~Billet,$^{1,2,3}$ Carlos~Mas~Arabi,$^{1}$ Fabrice~Raineri,$^{6,7}$ Sylvain~Combri\'e,$^8$ Alfredo~De~Rossi,$^8$  Gunther~Roelkens,$^{2,3}$ Simon-Pierre~Gorza,$^1$ Bart~Kuyken,$^{2,3}$ and Fran\c cois~Leo$^{1,*}$}

\affiliation{$^1$ OPERA-Photonique, Universit\'e libre de Bruxelles, Brussels, Belgium}
\affiliation{$^2$ Photonics Research Group, Ghent University-IMEC, Ghent, Belgium}
\affiliation{$^3$ Center for Nano- and Biophotonics (NB-Photonics), Ghent University, Ghent, Belgium}
\affiliation{$^4$ Department of Electrical Engineering, Columbia University, New York, USA}
\affiliation{$^5$ Laboratoire de Photonique d'Angers EA 4464, Universit\'e d'Angers , Angers, France}
\affiliation{$^6$ Centre de Nanosciences et de Nanotechnologies (C2N), CNRS, Univ. Paris Sud, Univ. Paris Saclay, F-91120 Palaiseau, France}
\affiliation{$^7$ Universit\'e de Paris, Centre de Nanosciences et de Nanotechnologies (C2N), F-91120 Palaiseau, France}
\affiliation{$^8$ Thales Research and Technology, Palaiseau, France}
\email{francois.leo@ulb.ac.be}

\begin{abstract}

We investigate type I second harmonic generation in III-V semiconductor wire waveguides aligned with a crystallographic axis. \textcolor{\textmodcolor}{In this direction, because of the single nonzero tensor element of III-V semiconductors, only frequency conversion by mixing with the longitudinal components of the optical fields is allowed}. We experimentally study the impact of the propagation direction on the conversion efficiency and confirm the role played by the longitudinal components through the excitation of an antisymmetric second harmonic higher order mode.

\end{abstract}
\maketitle

\section{\label{sec:level1} Introduction}

Nanophotonic wire waveguides are well suited for efficient frequency conversion. The high index contrast allows to confine waves into sub-wavelength structures, greatly increasing the strength of light-matter interactions. Recent examples include ultra-efficient second harmonic (SH) generation~\cite{wang_ultrahigh_2018,chang_heterogeneously_2018,Stanton_ShgGaAs_2020} and frequency comb formation~\cite{pu_efficient_2016,stern_battery_2018}.
The bound modes propagating in high index contrast waveguides are different from the well known transverse plane wave solutions of the Helmholtz equation~\cite{snyder_optical_1983}.
In particular, these modes possess a strong longitudinal component. In silicon wire waveguides for example, the amplitude of the longitudinal field can be almost as large as that of its transverse counterpart~\cite{driscoll_large_2009}. 
Having a high electric field pointing in the direction of propagation is of interest for applications such as linear accelerators~\cite{bochove_acceleration_1992}, optical storage~\cite{vandenes_high-density_2006} or microscopy~\cite{novotny_longitudinal_2001}. 
Yet, the strong longitudinal fields components in nanowaveguides are still to be harnessed for efficient nonlinear conversion.

\textcolor{\textmodcolor}{Considering an instantaneous response of the medium, the electric polarization is often written as~\cite{boyd_nonlinear_2003} $\vec{P}=\varepsilon_0(\chi^{(1)}\vert\vec{E}+\chi^{(2)}\vert\vec{E}\vec{E}+...)$, where $\vec{E}$ is the real electric field, $\chi^{(1)}$ and $\chi^{(2)}$ are second and third rank tensors modeling the linear and the second order nonlinear response of the medium, and $\vert$ denotes a tensorial multiplication. In nanowaveguides, the longitudinal field induces a linear coupling} between modes through the $\chi^{(1)}_{zz}$ term~\cite{vahid_full_2009}. Polarization rotators based on this principle have been implemented~\cite{dai_noverl_2011}. 
The impact of longitudinal components on the second order nonlinear polarization was studied in zinc selenide crystals, lithium niobate and gallium phosphide nanopillars~\cite{kozawa_observation_2008,baghban_impact_2016,deluca_modal_2017}
but, to the best of our knowledge, \textcolor{\textmodcolor}{the SH generation enabled by those components has never been demonstrated for bound modes.} 

\textcolor{\textmodcolor}{In this paper, we analyze type I second harmonic generation in indium gallium phosphide (InGaP) wire waveguides in the strong guidance regime. We first model the nonlinear interaction and predict that the longitudinal electric field components play a crucial role for specific propagation directions. We then experimentally investigate the nonlinear interaction and demonstrate how axially oscillating dipoles may be leveraged to excite an antisymmetric SH mode in III-V semiconductor wire waveguides, a process that is prohibited for purely transverse modes.} 

\section{\label{sec:level2} Modeling}

III-V semiconductors are arranged in a zinc blende configuration ($\bar{4}$3m point group) leaving only the nondiagonal $\chi^{(2)}_{xyz}$ element nonzero. It was measured to be as high as 220~pm/V in InGaP~\cite{ueno_second_1997}.
\textcolor{\textmodcolor}{Here, we assume that the waveguides are in the $xz$ (010) plane of the crystal. The Cartesian coordinates are denoted ($x'y'z'$) in the waveguide frame and ($x y z $) in the crystal axis.
Since we are interested in type I second harmonic generation, we write the electromagnetic radiation as a superposition of two forward propagating bound modes:}
\begin{eqnarray}\label{eqperturbedEfield}
\vec{E}=&\Re\{ a(z') {\vec{e}_{a}}e^{i( \beta_a z'-\omega_0 t)}+ b(z') {\vec{e}_b}e^{i( \beta_b z'-2\omega_0 t)}\},\nonumber\\
\vec{H}=&\Re\{ a(z') {\vec{h}_a}e^{i( \beta_a z'-\omega_0 t)}+b(z'){\vec{h}_b}e^{i( \beta_b z'-2\omega_0 t)}\}.
\end{eqnarray}
\textcolor{\textmodcolor}{where $a$, $b$ are the amplitudes of the fundamental and second harmonic modes (expressed in $\sqrt{\mathrm{W}}$), $\beta_a,\beta_b$ are their propagation constants at the carrier frequencies $\omega_0$ and $2\omega_0$ respectively. $\vec{e}_j(\vec{r'}_\perp,\omega_j)$ and $\vec{h}_j(\vec{r'}_\perp,\omega_j)$ are the vectorial electric and magnetic profile of the modes, normalized so that $\left|a\right|^2$ (resp. $\left|b\right|^2$) is the power carried by the fundamental (resp. SH) mode. Both the profiles and the propagation constants are computed with a commercial mode solvers (Lumerical).}

\begin{figure}[t]
\centering
 \includegraphics[width=8cm]{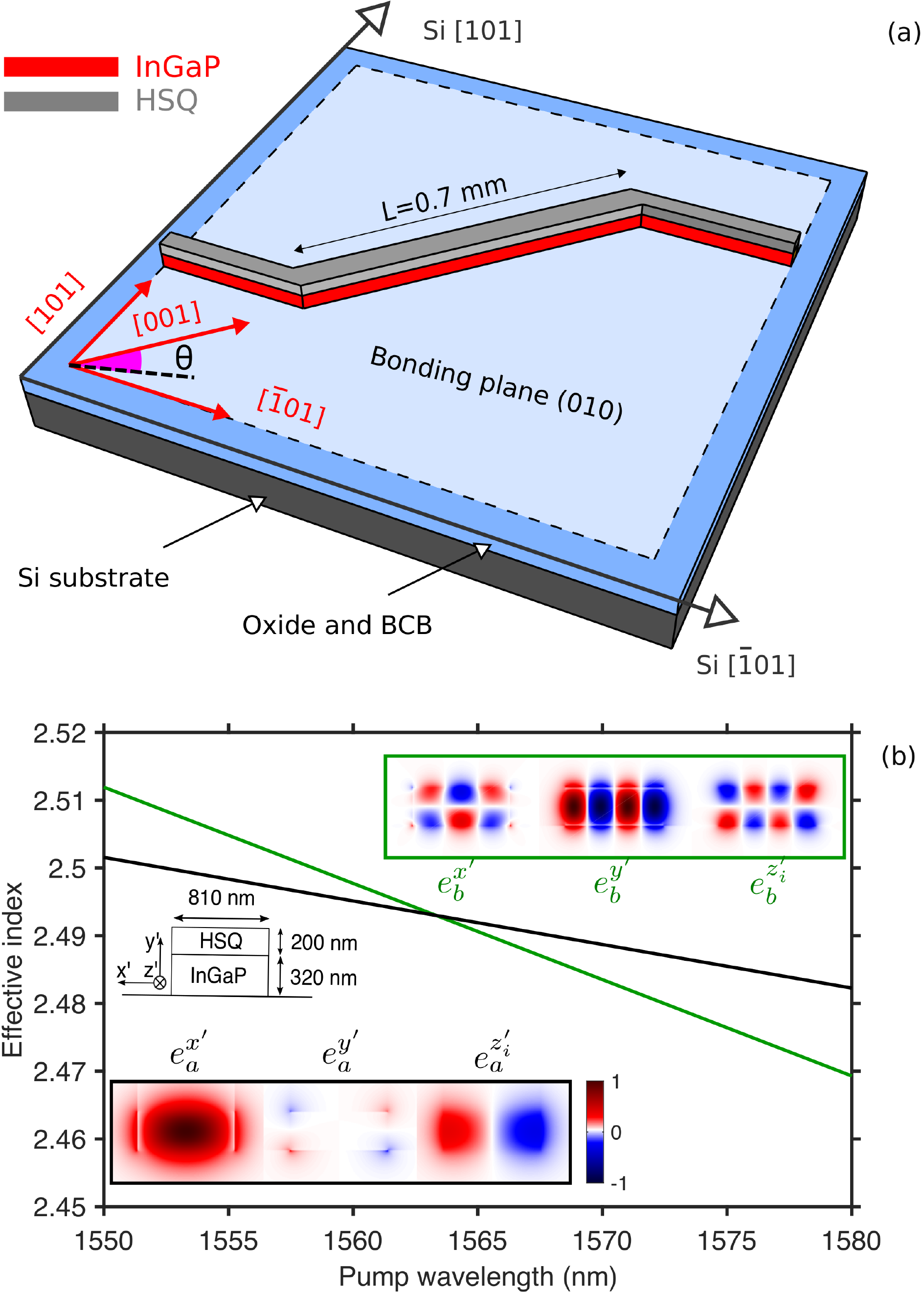}
\caption{(a) Schematic of a 320~nm high, 810~nm wide III-V-on insulator waveguide with both $\ang{0}$ and $\ang{45}$ sections.
(b) Simulated effective indices of the corresponding pump TE$_{00}$ and SH TM$_{30}$ modes showing phase matching close to 1560~nm. The waveguide cross section as well as the spatial profile of the electric field components are shown as insets.}
\label{figZ1}
\end{figure}

\textcolor{\textmodcolor}{The propagation model describing the nonlinear interaction between these modes can be derived by perturbation analysis of the linear bound modes~\cite{ciret_full_2019}. A similar approach has been applied to difference frequency generation~\cite{alloatti_second_2012} as well as third order nonlinearities~\cite{kolesik_nonlinear_2004,chen_theory_2006,vahid_full_2009,daniel_vectorial_2010,alexander_electrically_2017}. For type I second order interactions in waveguides, the nonlinear coupled equations read:}

\begin{eqnarray}
\frac{d a(z')}{d z'}&=&-\alpha_aa(z')+i\kappa^*(\theta)b(z')a^*(z')e^{-i\Delta\beta z'},\label{eq1kappa}\\
\frac{d b(z')}{d z'}&=&-\alpha_bb(z')+i\kappa(\theta) a^2(z')e^{i\Delta\beta z'}.
\label{eq2kappa}
\end{eqnarray}
where $\Delta\beta = 2\beta_a-\beta_b$ and $\theta$ is the angle between the propagation direction $z'$ \textcolor{\textmodcolor}{ and the $z$ axis ([001] direction, see Fig~\ref{figZ1}).} 

In the crystallographic axes, the effective nonlinearity is:
\begin{equation}\label{eqkappa}
\kappa = \frac{\omega_0\varepsilon_0}{2}\int_{\perp}\chi^{(2)}_{xyz}\left({e}^{*x}_b{e}^{y}_a{e}^{z}_a+{e}^{*y}_b{e}^{x}_a{e}^{z}_a +{e}^{*z}_b{e}^{x}_a{e}^{y}_a\right) dA,
\end{equation}
\textcolor{\textmodcolor}{where the integration boundaries are restricted to the transverse section of the InGaP waveguide core. We note }that the longitudinal components have a fixed phase difference of $\pi/2$ with their transverse counterparts. We choose to work with real transverse components such that the longitudinal components are purely imaginary. Accordingly, we introduce the imaginary part of the longitudinal component $ e^{z'_i}=-ie^{z'}$ which allows to use purely real spatial distributions in what follows. \textcolor{\textmodcolor}{Moreover we can limit the discussion to the first quadrant considering the $\bar{4}$ symmetry of the material.}

The conversion efficiency is either maximum at $\theta=\ang{0}$ or $\ang{45}$ depending on the spatial distributions of the modes~\cite{ciret_full_2019}. \textcolor{\textmodcolor}{When the waveguide is aligned with the [001] crystal axis, the effective nonlinearity becomes:}
\begin{align}\label{eqkappa0}
\kappa(\ang{0}) = \frac{i\omega_0\varepsilon_0}{2}\int_{\perp}\chi^{(2)}_{xyz}&\left({e}^{x'}_b{e}^{y'}_a{e}^{z'_i}_a+{e}^{y'}_b{e}^{x'}_a{e}^{z'_i}_a\right.\nonumber\\ &-\left. e^{z'_i}_be^{x'}_a{e}^{y'}_a\right)  dA.
\end{align}
\textcolor{\textmodcolor}{We readily see that in this particular propagation direction the longitudinal fields components play a critical role as they appear in all terms of Eq~\eqref{eqkappa0}.} 

\textcolor{\textmodcolor}{Conversely, when aligned at \ang{45}, the effective nonlinearity is very different since the transverse components can efficiently mix together while the longitudinal components contribution is usually small. It reads:} 
\begin{align}\label{eqkappa45}
\kappa(\ang{45}) = \frac{\omega_0\varepsilon_0}{2}\int_{\perp}\chi^{(2)}_{xyz}&\left[{e}^{y'}_a\left({e}^{x'}_b{e}^{x'}_a-{e}^{z'_i}_b{e}^{z'_i}_a\right)\right.\nonumber\\ &+\left. \frac{{e}^{y'}_b}{2}\left({e^{x'}_a}^2+{e^{z'_i}_a}^2\right)\right]  dA.
\end{align}

When a mirror symmetry is present, spatial distributions ${e}^i_{a,b}$ are either symmetric or antisymmetric in the corresponding direction [see e.g. Fig~\ref{figZ1} (b)]. \textcolor{\textmodcolor}{Moreover, } a symmetric (resp. antisymmetric) $e^x$ implies an antisymmetric (resp. symmetric) $e^y$  and $e^z$~\cite{snyder_optical_1983}. This has important consequences for the overlap integrals in Eqs~\eqref{eqkappa0} and \eqref{eqkappa45}.  Second harmonic modes with a certain symmetry can only be efficiently excited in a specific direction: the modes with a symmetric longitudinal profiles at $\ang{45}$ and the modes with an antisymmetric longitudinal profile at $\ang{0}$. Note that the symmetry of the pump mode has no impact because two pump photons are involved in the process. 
There are hence two experimental signatures of second harmonic generation induced by longitudinal components in \textcolor{\textmodcolor}{III-V}  wire waveguides: (i) the propagation direction that maximizes the conversion efficiency should be aligned with a crystallographic axis and (ii) in the $x'$ direction, the vertical and longitudinal components of the second harmonic mode, ($e^{y'}_b$,$e^{z'}_b$), are antisymmetric and the corresponding horizontal component ($e^{x'}_b$) is symmetric.

\section{\label{sec:level3} Experiments and discussion}

\begin{figure}[t]
\centering
 \includegraphics[width=8cm]{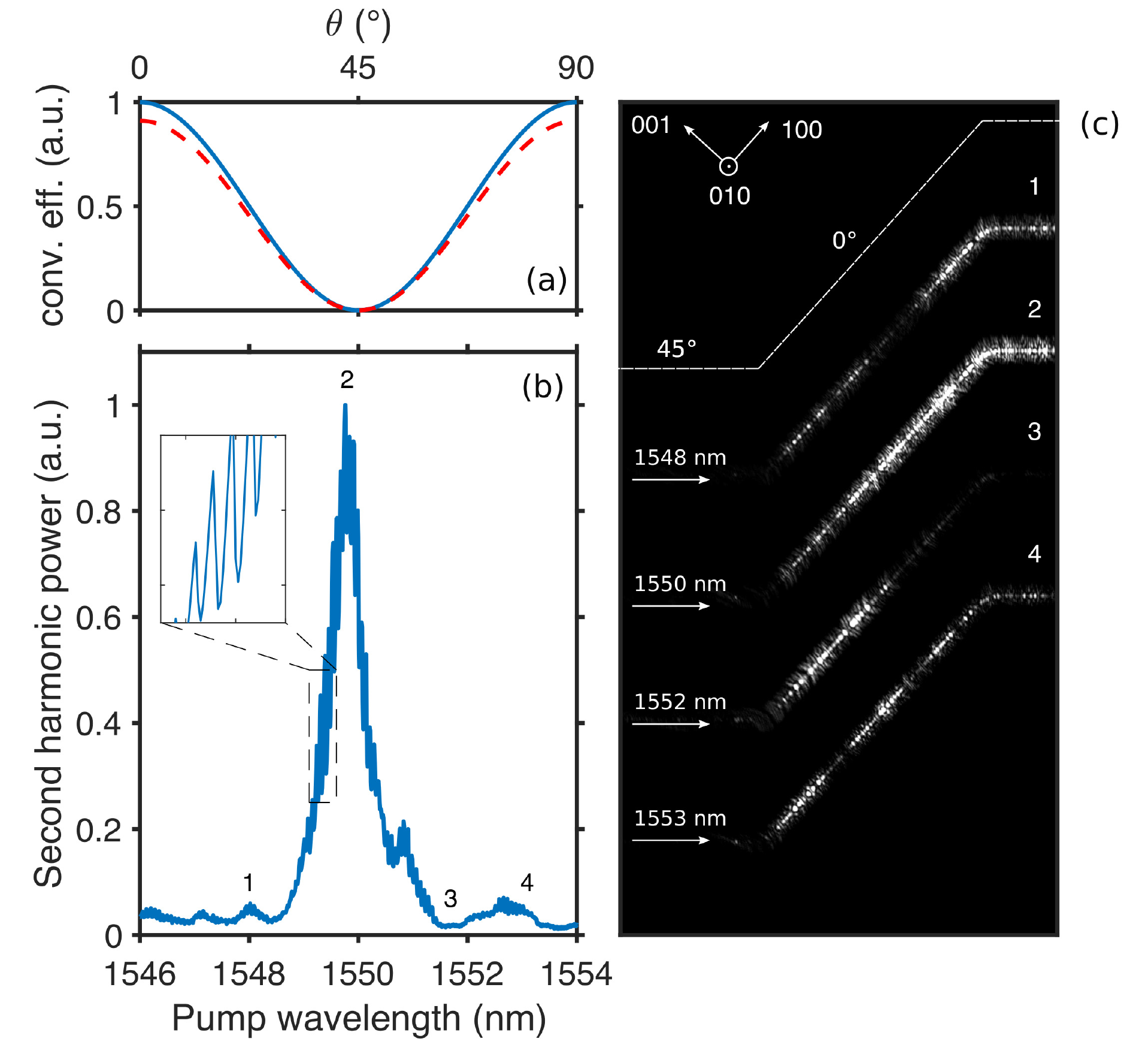}
\caption{(a) \textcolor{\textmodcolor}{Theoretical conversion efficiency ($|\kappa(\theta)|^2$) corresponding to the coupling shown in Fig~\ref{figZ1}, when considering all the terms of Eq~\eqref{eqkappa} (blue line) or only the ${e}^{y}_b{e}^{x}_a{e}^{z_i}_a$ term (red dashed line). (b) Measured} second harmonic power versus pump wavelength. A close up of the Fabry-Perot fringes (\textcolor{\textmodcolor}{ FSR = 75 pm}) is shown in the inset.  (c) \textcolor{\textmodcolor}{Top view of the diffusion patterns recorded on a silicon camera for different pump wavelengths (they are shifted vertically from one another for clarity).} The dashed line indicates the position of the $\ang{0}$ and $\ang{45}$ sections. The intensity has been normalized so as to show the patterns more clearly.}
\label{figZ2}
\end{figure}

We fabricate waveguides in an epitaxially grown indium gallium phosphide thin layer bonded on a silicon wafer~\cite{dave_nonlinear_2015}. We start with a waveguide made of both $\ang{0}$ and $\ang{45}$ sections [see Fig~\ref{figZ1}\,(a)]. The cleave directions of both silicon and InGaP are along the [101] and [$\bar{1}01$] axis, such that the middle section (L = 700~\micron) is aligned with a crystallographic axis. The waveguide is 320~nm high and 810~nm wide. 
Mode solver simulations predict phase matching between a quasi-TE$_{00}$ pump mode and a quasi-TM$_{30}$ second harmonic mode \textcolor{\textmodcolor}{close to 1560 nm [see Fig~\ref{figZ1}\,(b)]}. The dependence of the corresponding effective nonlinearity with the propagation direction is shown in Fig~\ref{figZ2}\,(a). Also shown is the amplitude of the middle term (${e}^{y}_b{e}^{x}_a{e}^{z_i}_a$) of Eq~\eqref{eqkappa} to highlight that the interaction between \textcolor{\textmodcolor}{these modes is dominated by the mixing of those components. At the phase matching wavelength, the maximum conversion corresponds to $\kappa(\ang{0})= 260i\mathrm{(\sqrt{W}m)^{-1}}$, while no conversion should occur at $\theta=\ang{45}$.}

We launch horizontally polarized light from a tunable laser in the waveguide through a lensed fiber. The output waves are collected by use of a microscope objective with a large numerical aperture (NA=0.9).  The experimental phase matching function is shown in Fig~\ref{figZ2}\,(b). We find maximum conversion close to 1550~nm with a conversion bandwidth of 0.6~nm.
The fast oscillations (see inset) are due to the Fabry-Perot effect induced by Fresnel reflections on both facets of the waveguide.
The measured \textcolor{\textmodcolor}{ 75~pm} free spectral range matches very well with the theoretical \textcolor{\textmodcolor}{ 78~pm} corresponding to a 4.4~mm long waveguide with a group index $n_g=3.5$. 
We also image the SH diffusion pattern from the top of the waveguide with a silicon camera for pump wavelengths corresponding to local extrema of the phase matching function [see Fig~\ref{figZ2}\,(c)]. 
\textcolor{\textmodcolor}{As can be seen, the SH wave appears shortly after the first turn and, at the phase matching wavelength, it increases until the $\ang{0}$ section ends.} This confirms our prediction that the nonlinear conversion occurs only in the waveguide aligned with the crystallographic axis.
Around 1548~nm we find a pattern of SH generation and parametric down conversion, in agreement with the dynamics for $\Delta\beta = \pm3\pi/L$~\cite{boyd_nonlinear_2003}.
When pumping at 1553~nm however, the measurement is reminiscent of the dynamics expected for $\Delta\beta = \pm5\pi/L$. This discrepancy is confirmed by pumping at 1552~nm  where nonlinear conversion seems to take place only in the first half of the $\ang{0}$ section.
The diffusion patterns \textcolor{\textmodcolor}{and the SH power curve} indicate that the conversion dynamics is different on both sides of the maximum conversion point, a behavior we could not replicate by integrating Eqs~\eqref{eq1kappa}-\eqref{eq2kappa}. 
Asymmetric phase matching functions have been \textcolor{\textmodcolor}{reported} in other SHG experiments in nanowires~\cite{duchesne_second_2011,luo_highly_2018,wang_ultrahigh_2018,chang_heterogeneously_2018,  Anthur_SHG-GAP_2020}. They are likely due to inhomogeneities of the waveguide cross section. 
A detailed study of these inhomogeneities on SH pattern is beyond the scope of the present report. 
Importantly, our experiments confirm the impact of the propagation direction on second harmonic generation in III-V nanowaveguides. The SH is most efficiently excited when the waveguide is aligned with a crystallographic axis, as theoretically predicted when the interaction is enabled by the longitudinal components of the modes.

\begin{figure}[t]
\centering
 \includegraphics[width=8cm]{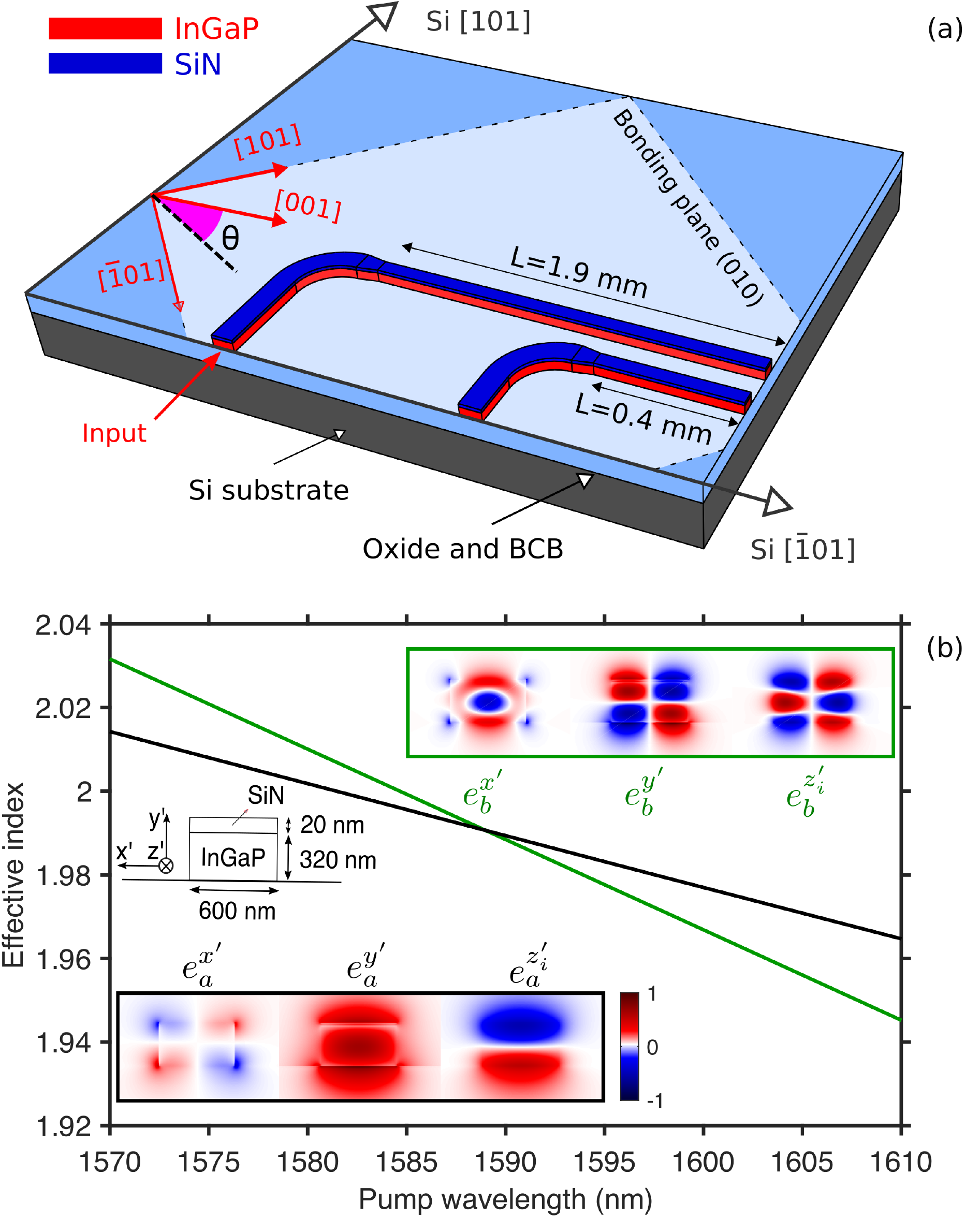}
\caption{(a) Schematic of two straight sections of different lengths of 320~nm high, 600~nm wide III-V-on insulator waveguides. \textcolor{\textmodcolor}{At the input}, both are connected to a 700~nm wide section.
(b) Simulated effective indices of the corresponding pump TM$_{00}$ and SH TM$_{11}$ modes showing phase matching around 1590~nm. The waveguide cross section as well as the spatial profile of the electric field components are shown as insets.}
\label{figL1}
\end{figure}

\begin{figure}[t]
\centering
 \includegraphics[width=8cm]{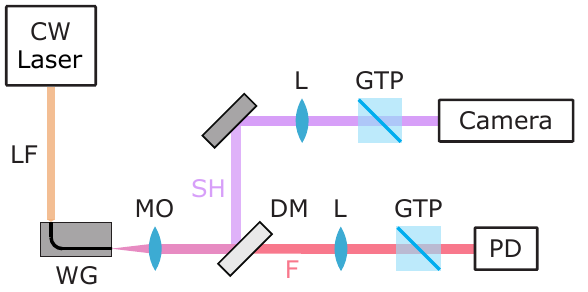}
\caption{Experimental setup. A lensed fibre (LF) is used to coupled the light in the waveguide (WG). At the output, a microscope objective (MO) collects the light, which is then sent a dichroic mirror (DM) that separates the fundamental (F) and SH beams. They are both collimated through a lens (L) and sent to a Glan-Taylor prism (GTP) and a photodiode (PD). Alternatively, as depicted in the figure, the SH mode is imaged on a camera.}
\label{figSetup}
\end{figure}

\textcolor{\textmodcolor}{In a second experiment, we aim at measuring the SH intensity profile to confirm the mixing with the longitudinal components of the fields.} To have access to the excited second harmonic mode, without disturbance from bends and sections with no SH conversion, we fabricate waveguides in an InGaP layer rotated with respect to the silicon wafer [see Fig~\ref{figL1}(a)]. \textcolor{\textmodcolor}{We resorted in this case} to a positive-tone resist (ARP-6200) for the lithography and silicon nitride as a hard mask.
We use waveguides with a $\ang{90}$ bend between sections of different width to be able to study different lengths while keeping the propagation direction fixed. 
\textcolor{\textmodcolor}{At the input, we couple linearly polarized light from a tunable laser source into both waveguides through the 700~nm wide section by use of a lensed fiber. 
The output modes are imaged through a 4-f arrangement comprising a high NA microscope objective (f = 1.8~mm) and a 750~mm plano-convex lens (see Fig~\ref{figSetup}). 
Because of the sub-wavelength nature of the transverse profile, only the lowest order modes can be faithfully imaged. The large point spread function of the imaging system} relative to the waveguide width prevents from reconstructing modes with more than one zero in \textcolor{\textmodcolor}{each} directions. \textcolor{\textmodcolor}{Hence, we seek to excite a SH antisymmetric mode of the lowest order in the x-direction. Constrained by the fixed InGaP layer height of 320~nm, we look for a mode with only one zero in this direction and one or no zero in the y-direction.}  Simulations predict phase matching between a quasi-TM$_{00}$ pump mode and a quasi-TM$_{11}$ second harmonic mode in a 600~nm wide waveguide for a pump around 1590~nm (see Fig~\ref{figL1}).
The corresponding theoretical effective nonlinearity is $\kappa(\ang{0})=-57i \mathrm{(\sqrt{W}m)^{-1}}$.
\textcolor{\textmodcolor}{Note that, contrary to the previous experiment, none of the three terms of Eq~\eqref{eqkappa0} are negligible. All field components, including both longitudinal ones, thus play a significant role in the nonlinear coupling. }

\begin{figure}[th]
\centering
 \includegraphics[width=8cm]{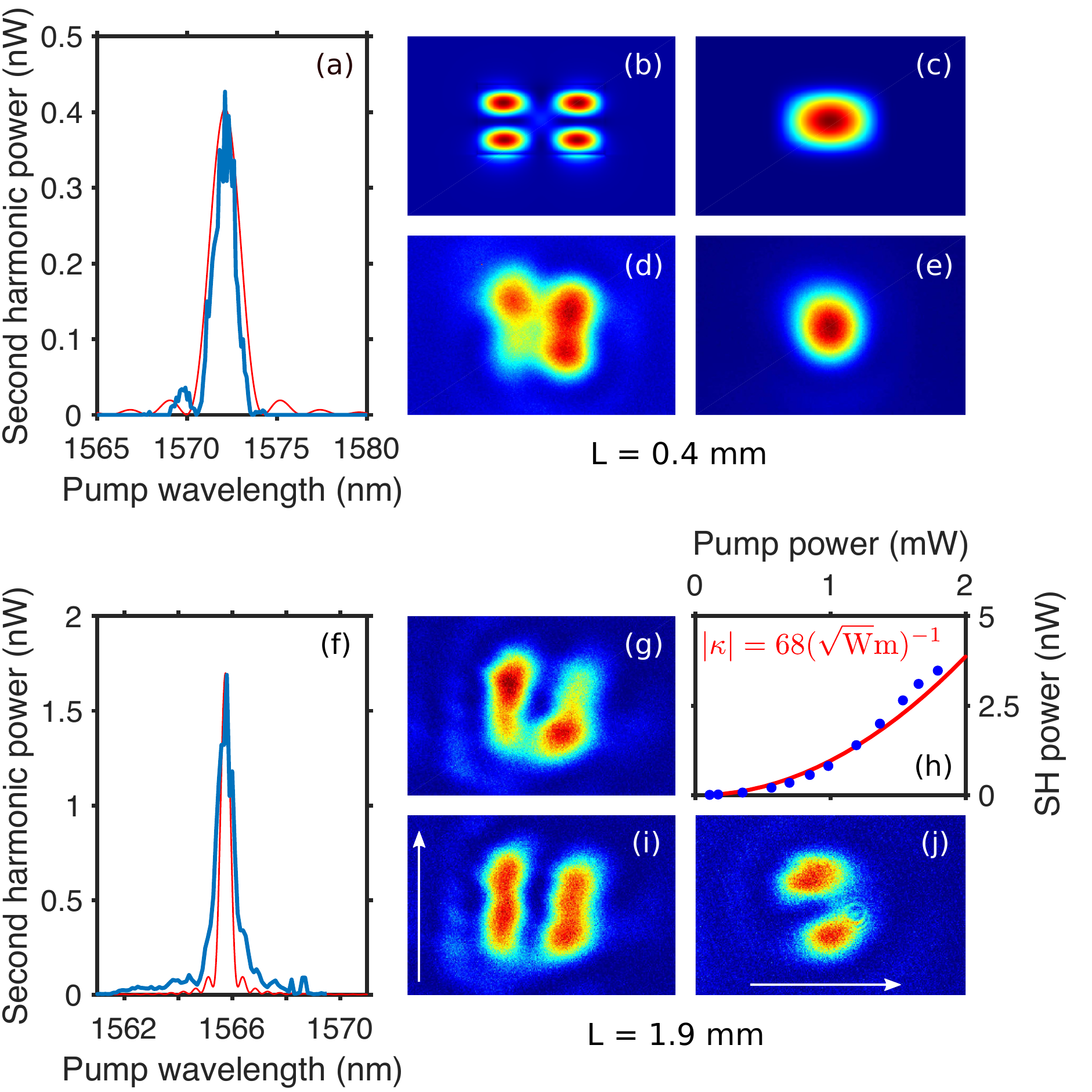}
\caption{Experimental results obtained with the waveguides shown in Fig~\ref{figL1}(a) Experimental (blue) and theoretical (red) SH harmonic power at the output of the short waveguide versus pump wavelength (input power = 2.5 mW). (b,c) Theoretical Poynting vector profile for the SH TM$_{11}$ mode and TE$_{00}$ mode. (d,e) Experimental profiles of the SH TM$_{11}$ mode and TE$_{00}$ mode. The theoretical and experimental field of views are 1.27~\micron~x~ 0.85~\micron~and 2.84~\micron~x~1.89~\micron~respectively. (f) Experimental (blue) and theoretical (red) SH power at the output of the long waveguide versus pump wavelength (input power = 1.3 mW). (g) Experimental image of the SH TM$_{11}$ mode at the output of the long waveguide. (h) Experimental (blue dots) and theoretical (red line) SH power as a function of the pump power when pumped at 1566~nm. (i,j) Experimental image of the SH mode after the Glan-Taylor prism in the vertical and horizontal directions, as indicated by the white arrow.}
\label{figL2}
\end{figure}

The maximum conversion is experimentally found at 1572~nm in the shorter waveguide (see Fig~\ref{figL2}). \textcolor{\textmodcolor}{The analysis of the polarization of the output modes through a Glan-Taylor prism shows that both the pump and the SH waves are predominantly vertically polarized, as predicted by the full-vectorial modeling.}

\textcolor{\textmodcolor}{To calibrate our imaging system at the second harmonic wavelength, we launch 775~nm horizontally polarized light at the input of the waveguide to excite a fundamental mode}. We measure a profile very similar to the computed TE$_{00}$ mode [Fig~\ref{figL2}(c,e)] given the limited resolution of the imaging setup. 
We then image the output SH mode when pumping at 1572~nm and find \textcolor{\textmodcolor}{very good} agreement between theoretical and experimental profiles [Fig~\ref{figL2}(b,d)]. 
In the longer waveguide, we find a similar phase matching point around 1566~nm. The phase matching function is narrower, as expected, and the output mode profile confirms that the same TM$_{11}$ SH mode as in the short waveguide is excited [Fig~\ref{figL2}(g)]. 
As discussed above, different symmetries are expected for the different components. The vertically (resp. horizontally) polarized component should be antisymmetric (resp. symmetric) in the $x'$ direction. \textcolor{\textmodcolor}{To find out, we record the image profile through a Glan-Taylor prism. The horizontally and vertically polarized profiles at the output of the 1.9~mm waveguide} are shown in Fig~\ref{figL2}(i,j). The vertically polarized wave is very similar to the Poyting vector distribution, as expected from a quasi-TM mode. After the prism though, a clearer zero line can be seen along the symmetry plane. The horizontal polarization pattern on the other hand is very different. While not perfectly symmetric as would be expected, it is clear that there is no vertical nodal line in the intensity profile. Note that both plots are normalized \textcolor{\textmodcolor}{and that the power in the horizontally polarized component} is around 10\% of the total output power.
\textcolor{\textmodcolor}{These results clearly confirm the excitation of a TM$_{11}$ SH mode by a TM$_{00}$ pump mode in a III-V wire waveguide. Since this process is only supported by the mixing of all six electric field components, it highlights the importance of the longitudinal field components and the necessity to perform full-vectorial modeling when designing SHG experiments in III-V nanowires.}

\textcolor{\textmodcolor}{Finally, we extract an experimental lower bound to the effective nonlinear coefficient from the conversion efficiency. The theoretical efficiency can be obtained by integrating Eqs~\eqref{eq1kappa}-\eqref{eq2kappa} while neglecting pump depletion \cite{Pliska98_SHGwaveguideLiNbO}:} 
\begin{align}\label{eqpm}
\left|b(L)\right|^2 =4|\kappa|^2\left|a(0)\right|^4\left| \frac{\mathrm{sinh}(\Phi L/2)}{\Phi}\right|^2e^{-(\alpha_a+\alpha_b/2)L},
\end{align} 
\textcolor{\textmodcolor}{where $\Phi = \alpha_a -\alpha_b/2 +i\Delta\beta$. To extract the experimental efficiency from the SH power measurements,} [see Fig~\ref{figL2}(h)], we use the theoretical maximum value $\left|b(L)\right|^2 =  |\kappa|^2\left|a(0)\right|^4L^2_{\mathrm{eff}}$, where we have introduced the effective interaction length $L_{\mathrm{eff}}=2[\exp{(- \alpha_bL/2)}-\exp{(-\alpha_aL)}]/(2\alpha_a-\alpha_b)$. The input coupling and propagation loss of the pump are estimated to be 12~dB and 2.2~dB/mm respectively. The 775~nm propagation loss (9~dB/mm) was measured for a fundamental TE$_{00}$ at 775~nm. We expect the propagation loss of the TM$_{11}$ mode to be larger. Considering that we collect all the SH power through the high NA objective, we find $\left|\kappa_{\mathrm{exp}}\right| > 27\; \mathrm{(\sqrt{W}m)^{-1}}$ for the short waveguide and $\left|\kappa_{\mathrm{exp}}\right| > 68\; \mathrm{(\sqrt{W}m)^{-1}}$ for the long waveguide, in \textcolor{\textmodcolor}{good agreement with the theoretically predicted value of $57\; \mathrm{(\sqrt{W}m)^{-1}}$. Note that, as seen in Fig~\ref{figL2}(a,f), the spectral dependence of the process still resembles the standard sine cardinal squared function despite the large propagation loss, and the experimental bandwidth matches very well with the theoretical one in both waveguides.} 

\section{\label{sec:level4} Conclusion}

In conclusion, we designed two experiments to investigate the impact of longitudinal fields components on second harmonic generation in III-V nanowires. We have demonstrated, through characterization of the propagation direction, spatial profiles, polarization and efficiency, the excitation of anti-symmetric modes via type I second harmonic generation. We theoretically showed that such coupling is enabled by the longitudinal components of the modes. Integrated platforms are likely to play a large role in future optical systems. A good understanding of the physics of confined modes is hence crucial in developing all-optical computing. Our results show that, while they do not contribute to the time-averaged Poynting vector, longitudinal components store energy that can be harnessed for nonlinear conversion. With further dispersion engineering, this scheme can lead to ultra-efficient second harmonic generation~\cite{ciret_full_2019}, a process that finds applications in many fields, including quantum circuits~\cite{zaske_visible_2012}.

\section*{Acknowledgements}
We would like to thank Pascal Kockaert for fruitful discussions.   
This work was supported by funding from the European Research Council (ERC) under the European Union’s Horizon 2020 research and innovation programme (grant agreement Nos 726420,  759483 \& 757800) and by the Fonds de la Recherche Fondamentale Collective (grant agreement No PDR.T.0185.18).

\end{document}